# Fast, Responsive Decentralised Graph Colouring

Alessandro Checco and Doug J. Leith

*Abstract*—Graph colouring problem arises in numerous networking applications. We solve it in a fully decentralised way (*i.e.* with no message passing). We propose a novel algorithm that is automatically responsive to topology changes, and we prove that it converges to a proper colouring in $\mathcal{O}(N \log N)$ time with high probability for generic graphs when the number of available colours is greater than $\Delta$, the maximum degree of the graph, and in $\mathcal{O}(\log N)$ time if $\Delta = \mathcal{O}(1)$. We believe the proof techniques used in this work are of independent interest and provide new insight into the properties required to ensure fast convergence of decentralised algorithms.

## I. INTRODUCTION

Many fundamental wireless network allocation tasks can be formulated as a colouring problem, including channel and sub-carrier allocation [8], TDMA scheduling [1, 9], scrambling code allocation [6], network coding [8] and so on. Importantly, these tasks must often be solved while respecting strong communication constraints due, for example, to the range over which devices can communicate being smaller than the range over which they interfere or otherwise interact. Moreover, the network topology can change over time, requiring the nodes to dynamically adapt to this. Recently, fully decentralised Communication-Free Learning (CFL) algorithms have been proposed for solving general constraint satisfaction problems without the need for message-passing [8] and with the ability to respond automatically to topology changes. These Communication-Free Learning (CFL) algorithms exploit local sensing to infer satisfaction/dissatisfaction of constraints, thereby avoiding the need for message-passing and use stochastic learning to converge to a satisfying assignment. The extension to colouring problems with strong sensing restrictions is investigated in [5].

However, this class of algorithms has not been proven to be fast: the only analytic bound available is exponential in the number of nodes. Further, even though the performance observed in simulations is typically sub-exponential, the dependency of the convergence rate on the algorithm parameters is not clear.

In this paper we address both of these issues. We propose a novel algorithm that is provably fast ($\mathcal{O}(N \log N)$) when the number of available colours is greater than $\Delta$ (the maximum degree of the graph). Moreover, this algorithm is well suited to implementation on resource constrained devices such as RFID tags, because it has a small memory and computation footprint, involves no floating point arithmetic, no multiplications or divisions and only needs the availability of a uniform random number generator. The algorithm makes use of a new type of "memory" when learning a colouring and this gives some insight into which parts of the existing stochastic learning algorithms are needed to ensure fast convergence. We analyse how the algorithm parameters affect the convergence rate, and determine choices that guarantee fast convergence while maintaining responsiveness to topology changes.

Our theoretical bounds on convergence rate are obtained using a novel stochastic drift analysis that we believe is of independent interest.

### A. Related Work

In the graph theory and computer science literature, the problem of colouring with $\Delta + 1$ colours has been thoroughly studied [12, 15, 17, 21]. In particular, the family of *locally iterative algorithms* has received much attention. This family of algorithms makes use of the following strong assumptions:

1) The algorithm can use an unbounded number of colours during its operation, although it eventually reduces the number over time;
2) The graph topology is assumed to be known and fixed;
3) Each graph vertex needs to know which colours are used by its neighbours.

Szegedy and Vishwanathan [21] use an heuristic argument to show that no locally iterative $(\Delta + 1)$-colouring algorithm is likely to terminate in less than $\Omega(\Delta \log \Delta)$ rounds (lower bound). More refined bounds have been obtained in [3, 11].

However, in the wireless networking field, these three assumptions are rarely satisfied. To our knowledge, Assumption 1 is not used by any work in the wireless networking field, presumably because it is obviously inappropriate for such applications. Assumption 2 has been relaxed in two main ways: by using network-wide stopping/restarting techniques in annealing-like algorithms [13], and by use of learning algorithms [2, 4, 8, 14, 15, 21]. Assumption 3 (in the form of either centralised or gossiping-like message passing) is used in [7, 12, 13, 20, 23]. However, communication between nodes often cannot be relied upon in the design of a robust algorithm in cases where *e.g.* wireless nodes belong to different administrative domains or when the devices are too simple to be able to realise such communication (see, for example, Radio-Frequency Identification devices).

The most challenging problem of graph colouring in which no message passing is possible (so none of the three assumptions hold) has recently attracted attention in the wireless networking literature in [2, 8, 18].

The Learning-BEB algorithm, proposed by Barcelo et al. [2] is an algorithm devised for achieving collision-free scheduling in 802.11 networks. It is a modification of the CSMA/CA

A. Checco is with the Information School, University of Sheffield, UK. Email: a.checco@sheffield.ac.uk.

D.J. Leith is with the School of Computer Science and Statistics, Trinity College Dublin, Ireland. Email: doug.leith@scss.tcd.ie

This work was supported by Science Foundation Ireland under Grant No. 11/PI/1177.

Manuscript received October 27, 2015; revised September 1, 2017.

mechanism of truncated exponential backoff: after a successful transmission, the transmitter uses a fixed backoff interval $P$, while after a collision it selects an interval uniformly at random (u.a.r.) in the contention window range. Within the terminology of graph theory, this corresponds to a colouring algorithm in which each node selects the same colour after being locally satisfied, and selects a colour u.a.r. otherwise. This algorithm is known to suffer from slow convergence rates [9], but it has the advantage of being easy to implement. Roughly speaking, for a complete graph the probability of success for the whole graph in a step is $(1/N)^N$. So, the number of iterations needed to completely colour a graph with this method with probability bigger than $1 - \epsilon$ is of the order of $log(\epsilon^{-1})e^{(NlogN)}$, that is exponential in $N$.

The algorithm proposed by Motskin et al. [18] is similar to [2], with the advantage of being provably fast ($\mathcal{O}(\log N)$ when $\Delta = o(N)$) but with the major disadvantage of not being adaptive to topology changes, since after a correct local choice, the node keeps the chosen colour forever. The possibility of change in topology, together with the constraint of no message passing, makes this algorithm unable to "reset" when a change of topology happens, and thus this intuitive approach does not work.

The CFL algorithm proposed by Duffy et al. [8] uses a stochastic learning mechanism to update the probability of choosing each colour based on local sensing. In simulations it is fast, and it is provably adaptive to topology changes. The main disadvantage is that it is hard to prove good convergence rate bounds and it is too complicated to implement in simple hardware such as Radio-Frequency Identification (RFID) tags.

It is worth noting that all three algorithms share the common property of initially selecting colours u.a.r. and staying with the same colour when locally satisfied. They also all belong to the family of locally iterative algorithm, even if they do not use assumptions 1-3. The difference between them lies in the way they respond to a loss of local satisfaction: Learning-BEB will go back to u.a.r. selection, the algorithm proposed by Motskin et al. [18] will keep the same choice even if locally unsatisfied, and CFL will distribute the probability mass amongst all colours, decreasing the probability of choosing the current unsatisfying colour. Learning-BEB is equivalent to CFL when the latter uses parameters $a = b = 1$ (in the terminology of [8]).

To summarise, when message passing is not allowed and the graph topology can change over time, the existing solutions cannot provide a provably fast colouring, because of one of these reasons:

- The ability to react to a topology change creates convergence speed problems (Learning-BEB [2]).
- The ability to quickly converge comes at the expense of being unable to react to topology changes (algorithm in [18]).
- The ability to learn from the local graph topology by passive observation of the local state of a vertex via stochastic mechanisms make very arduous to prove fast convergence (CFL [8]).

Our approach allows to gain some insight into which parts of the existing stochastic learning algorithms are needed to ensure fast convergence, so that a simple and provably fast algorithm can be devised.

## II. PRELIMINARIES: PROBLEM DEFINITION

We use the notation introduced in [8] for the more general decentralised constraint satisfaction problem, applying it to graph colouring problem as in [5].

Let $G = (\mathcal{N}, \mathcal{E})$ denote an undirected graph with set of vertices $\mathcal{N} = \{1, \ldots, N\}$ and set of edges $\mathcal{E} := \{(i,j) : i, j \in \mathcal{N}, i \leftrightarrow j\}$, where $i \leftrightarrow j$ denotes the existence of an edge between $i$ and $j$. Let $\Delta$ denote the maximum degree of vertices in graph $G$ i.e. the maximum number of neighbours. A Colouring Problem (CP) on graph $G$ with $D \in \mathbb{N}$ colours is defined as follows. Let $x_i \in \mathcal{D}$ denote the colour of vertex $i$, where $\mathcal{D} = \{1, \ldots, D\}$ is the set of available colours, and $\vec{x}$ denote the vector $(x_1, \ldots, x_N)$. Define clause $\Phi_m \colon \mathcal{D}^N \mapsto \{0,1\}$ for each edge $m = (i,j) \in \mathcal{E}$ with:

$$\Phi_m(\vec{x}) = \Phi_m(x_i, x_j) = \begin{cases} 1 & \text{if } x_i \neq x_j \\ 0 & \text{otherwise.} \end{cases}$$

We say clause $\Phi_m(\vec{x})$ is *satisfied* if $\Phi_m(\vec{x}) = 1$. An assignment $\vec{x}$ is said to be satisfying if for all edges $m \in \mathcal{E}$ we have $\Phi_m(\vec{x}) = 1$. That is

$$\vec{x} \text{ is a satisfying assignment iff } \min_{m \in \mathcal{E}} \Phi_m(\vec{x}) = 1. \quad (1)$$

Equivalently, $\vec{x}$ is a satisfying assignment if and only if $x_i \neq x_j$ for all edges $(i,j) \in \mathcal{E}$ i.e. if $i \leftrightarrow j$. For any colour allocation, we say a vertex is *unsatisfied* if at least one of its neighbours has the same colour; otherwise the vertex is said to be *satisfied*. A satisfying assignment for a colouring problem is also called a *proper colouring*.

**Definition 1** (Chromatic Number)**.** The chromatic number $\chi(G)$ of graph $G$ is the smallest number of colours such that at least one proper colouring of $G$ exists.

We require the number of colours $D$ in our palette to be greater than or equal to $\chi(G)$ for a satisfying assignment to exist.

### A. Decentralized CP Solvers

**Definition 2** (CP solver)**.** Given a CP, a CP solver realizes a sequence of vectors $\{\vec{x}(t)\}$ such that for any CP that has a satisfying assignment

**(D1)** for all $t$ sufficiently large $\vec{x}(t) = \vec{x}$ for some satisfying assignment $\vec{x}$;

**(D2)** if $t'$ is the first entry in the sequence $\{\vec{x}(t)\}$ such that $\vec{x}(t')$ is a satisfying assignment, then $\vec{x}(t) = \vec{x}(t')$ for all $t > t'$.

In order to give criteria for classification of decentralized CP solvers, we re-write the LHS of Equation (1) to focus on the satisfaction of each variable

$$\vec{x} \text{ is a satisfying assignment iff } \min_{i \in \mathcal{N}} \min_{m \in \mathcal{E}_i} \Phi_m(\vec{x}) = 1. \quad (2)$$

where $\mathcal{E}_i$ consists of all edges in $\mathcal{E}$ that contain vertex $i$, i.e.

$$\mathcal{E}_i = \{(j,i) \colon (j,i) \in \mathcal{E}\}.$$

A decentralized CP solver is equivalent to a parallel solver, where each variable $x_i$ runs independently an instance of the solver, having only the information on whether all of the clauses that $x_i$ participates in are satisfied or at least one clause is unsatisfied. The solver located at variable $x_i$ must make its decisions only relying on this information.

**Definition 3** (Decentralized CP solver). A decentralized CP solver is a CP solver that for each variable $x_i$, must select its next value based only on the evaluation of

$$\min_{m \in \mathcal{E}_i} \Phi_m(\vec{x}). \tag{3}$$

That is, the decision is made without knowing

**(D3)** the assignment of $x_j$ for $j \neq i$.
**(D4)** the set of clauses that any variable, including itself, participates in, $\mathcal{E}_j$ for $j \in \mathcal{N}$.
**(D5)** the clauses $\Phi_m$ for $m \in \mathcal{E}$.

## III. FAST COLOURING

The basic idea used in [8] to establish convergence is to show that (i) for any graph and starting from any choice of colours there exists a sample path that leads to a proper colouring and (ii) over $\mathcal{O}(N)$ iterations this sample path occurs with probability bounded away from zero. While providing a bound on the convergence rate that appears to be order optimal, this approach cannot be used to show that fast convergence occurs when the number $D$ of colours is at least $\Delta + 1$. For that, we need an approach that provides insight into how the fraction of sample paths leading to a proper colouring *increases* with $D$. In this paper we therefore adopt an entirely different approach from that in [8], namely one based on stochastic drift analysis.

### A. FCFL Algorithm

As will become clear, it will prove helpful to consider the following generalisation of the CFL algorithm, called Fast Communication-Free Learning (FCFL):

---

**Algorithm 1** Fast Communication-Free Learning

1: Define $\mathcal{S}_\tau \in \mathbb{N}$, $\tau = 1, 2, \ldots$ with $S_{\tau+1} \geq S_\tau$
2: Define $D$, the number of colours, parameter $0 < b \leq 1$ and vector $\mathbf{p} \in [0,1]^D$
3: Initialise $\mathbf{p} = \frac{1}{D}\mathbf{1}$, $m = 0$, counters $\tau = 1$, $t = 1$
4: **repeat**
5:   **if** $t = S_\tau$ **then**
6:     $\tau = \tau + 1$, $m = 0$   ▷ Reset, exit permanent state
7:   **end if**
8:   Select colour $c$ with probability $p_c$
9:   **if** $m = 0$ **then**
10:     **if** Satisfied **then**
11:       $\mathbf{p} = \delta_c$   ▷ Stick with same colour
12:       $m = 1$   ▷ Enter permanent state
13:     **else**
14:       $\mathbf{p} = (1-b)\mathbf{p} + \frac{b}{D}\mathbf{1}$   ▷ Unsatisfied
15:     **end if**
16:   **end if**
17:   $t = t + 1$
18: **until** Forever

---

Each vertex maintains a vector $\mathbf{p} \in [0,1]^D$, where $D$ is the number of available colours, and a state $m \in \{0,1\}$. When $m = 1$ the vertex is said to be in the *permanent* state. Time is slotted with slots indexed by $t = 1, 2, \cdots$. A copy of the FCFL algorithm is run by every vertex, which determines vector $\mathbf{p}$ and state $m$ at each time slot. State $m$ is initially 0, and is also reset to 0 at times $\mathcal{S}_\tau$, $\tau = 1, 2, \ldots$. When $m = 0$, at each time slot a vertex selects a colour $c$ with probability $p_c$ (the $c$'th element of vector $\mathbf{p}$). If none of its neighbours have selected the same colour, the vertex is said to be *satisfied* and (i) $\mathbf{p}$ is updated so that all elements are 0 except for element $p_c$ which is set equal to 1, (ii) the vertex enters the permanent state $m = 1$. The vertex will therefore continue to select colour $c$ until the next reset time[1] $S_\tau$. However, if one or more neighbours have selected the same colour then the vertex updates $\mathbf{p}$ to $(1-b)\mathbf{p} + \frac{b}{D}\mathbf{1}$, reducing the probability with with colour $c$ is chosen, and tries again.

Evidently FCFL satisfies requirements D2-D5 of a CP solver, and we consider requirement D1 shortly. FCFL includes CFL and the related algorithms in [2, 18] as special cases. When $S_\tau = \tau$, so that the state $m$ is reset to 0 at every time step, then FCFL is equivalent to CFL[2]. When, in addition, $b = 1$ then unsatisfied vertices select colours uniformly at random ($\mathbf{p}$ equals $\frac{1}{D}\mathbf{1}$) and FCFL is equivalent to that considered in [2]. When $S_1 \to \infty$ and $b = 1$ then FCFL is equivalent to that in [18] (unsatisfied vertices select colours uniformly at random and once satisfied a vertex then sticks with the same colour forever).

Intuitively, parameter $b$ affects the memory in the algorithm. For example, suppose a vertex has selected colour $c$ and been satisfied so that $p_c = 1$ but has subsequently become unsatisfied (one of its neighbours has selected colour $c$). When $b$ is small, then $p_c$ reduces only slowly and so the vertex will tend to retry colour $c$ repeatedly. Conversely, when $b = 1$ then $p_c$ is immediately reset to $1/D$ and becomes as equally likely as any other colour to be selected at the next time slot. The interval $S_{\tau+1} - S_\tau$ between reset times similarly affects the rate of adaptation. When this interval is large then a vertex that has been satisfied keeps trying the same colour even when it becomes unsatisfied, and conversely when the interval is small. We will discuss the impact of $b$ and $S_\tau$ on the convergence rate in more detail below.

### B. Absorbing State and Dynamic Recolouring

Observe that the FCFL algorithm involves no stopping/restarting but a proper colouring is an absorbing state for FCFL, corresponding to a situation where all vertices are satisfied *i.e.* once all vertices are satisfied then their colours will remain fixed. Importantly, the situation where all vertices are in the permanent state is also absorbing:

---

[1]Note that the permanent state differs from the satisfied and unsatisfied states because a permanent vertex is guaranteed to be satisfied only at the round in which it becomes permanent.

[2]We found that the key difficulty with establishing the fast convergence of CFL is that the algorithm potentially has infinite memory (after a "success" the probability assigned to a colour is only gradually reduced upon "failure", except when there is "success" on another colour). In FCFL, when $b = 1$ the probability is guaranteed to be reset after at most $M$ failures and in this sense has only finite memory.

**Lemma 1** (Absorbing States). *If all vertices are in the permanent state, then they are all satisfied.*

*Proof:* First let us note that in the first round in which a vertex becomes permanent, it is satisfied and it cannot cause dissatisfaction to its neighbours; the neighbours can still be unsatisfied, but only because of other vertices. By contradiction, assume all vertices are in permanent state but there is at least one vertex $i$ unsatisfied. So there must be at least another neighbour $j$ unsatisfied and with same colour of $i$, by symmetry of dissatisfaction sensing. Now let us call $t_i, t_j$ the (last) time in which $i$ and $j$ became permanent, respectively. Assume, w.l.o.g. that $t_i < t_j$ (note that equality is not possible, because at first round a vertex becomes permanent it is necessarily satisfied). Now at time $t_j$, $j$ became permanent, so it chose a colour different from $i$, causing a contradiction. ∎

It will also prove useful later to note the following:

**Corollary 1.** *A vertex in the permanent state can be unsatisfied only by non-permanent neighbours.*

Observe also that a useful feature of the FCFL algorithm (and of CFL for that matter) is that should the graph subsequently change, for example upon the appearance of a new vertex, such that the colouring is no longer proper then one or more vertices are not satisfied and these will automatically restart searching for a proper colouring after the next reset time $S_{\tau+1}$ has been reached. That is, there is no need for a co-ordinated restart with its associated communication overhead.

### C. Convergence

Basic convergence of the FCFL algorithm can be shown an approach similar to that used in [8] for the CFL algorithm. Namely, we have:

**Theorem 1** (Convergence). *Consider a feasible CP on a graph $G = \{\mathcal{N}, \mathcal{E}\}$, with palette $\mathcal{D}$. Suppose the FCFL reset times satisfy $S_{\tau+1} - S_\tau \leq M$, $\tau = 1, 2, \cdots$ and $S_1 \leq M$. Given any unsatisfied assignment of colours $\vec{x}(0) \in \mathcal{D}^N$, then with probability greater than $1 - \epsilon \in (0, 1)$, the number of iterations for the FCFL algorithm to find a proper colouring is less than*

$$MN \exp(\tfrac{MN(N+1)}{2} \log(D)) \log(\epsilon^{-1}).$$

*Proof:* See Appendix. ∎

However, while we know that FCFL converges, the bound on the convergence rate in Theorem 1 is not sufficient to establish fast convergence when the number of colours is $\Delta + 1$ or greater.

### D. Outline of Drift-based Analysis

Let the random variable $Z_t$ denote the number of vertices which are *not* in the permanent state at time $t$. Recall that, by Lemma 1, $Z_t = 0$ is an absorbing state. Letting $R \in \{1, 2, \cdots\}$ be the earliest time such that $Z_R = 0$. Since $Z_t$ is integer valued it follows that $\mathbb{P}(R > t) = \mathbb{P}(Z_t \geq 1)$ and by Markov's inequality,

$$0 \leq \mathbb{P}(Z_t \geq 1) \leq \mathbb{E}[Z_t].$$

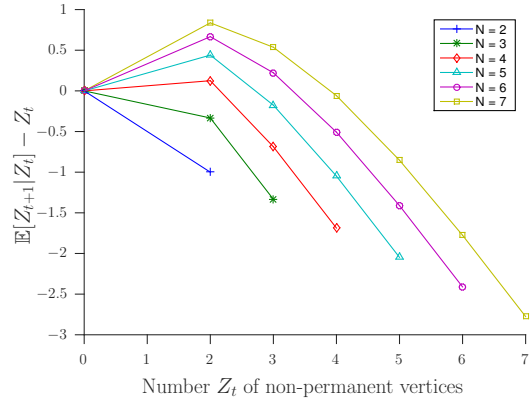

Fig. 1: Expected change in number of non-permanent vertices vs current number of non-permanent vertices. Results are shown for a complete graph with $N$ vertices, $D = \Delta + 1 = N$ colours and the FCFL algorithm with parameters $S_\tau = \tau$, $b = 1$.

Hence, it is sufficient to show that $\mathbb{E}[Z_t] \to 0$ in order to ensure that $\mathbb{P}(Z_t \geq 1) \to 0$ and so $\mathbb{P}(Z_t = 0) \to 1$. Further the convergence time of $\mathbb{E}[Z_t]$ upper bounds the convergence time of $\mathbb{P}(Z_t \geq 1)$.

If we could show that at all times $t$ the drift $\mathbb{E}[Z_{t+1} | Z_t] - Z_t \leq -\epsilon$ for some $\epsilon > 0$ then this would be enough to ensure that $\mathbb{E}[Z_t]$ decreases monotonically with time and also allow the convergence rate to be upper bounded. Indeed, this is the approach taken in [18] in the special case of FCFL where $S_1 \to \infty$.

Unfortunately, this approach cannot be used more generally since when the reset times $S_\tau$ are finite then the drift may, in fact, be positive at these times *i.e.* the number of vertices in the permanent state can decrease. Further, this positive drift is essential to ensure that the algorithm is able to respond to changes (such as addition of new vertices) in the graph which require recolouring of vertices. Figure 1 illustrates this, showing the expected change in the number of non-permanent vertices vs the current number of non-permanent vertices for a complete graph (each vertex is connected to every other vertex, so the degree $\Delta$ is $N - 1$). For $N > 3$ it can be seen that the drift is positive as the number of non-permanent vertices becomes small. This is quite intuitive: suppose two vertices are in the non-permanent state and so unsatisfied. Only in the relatively unlikely event in which they choose the remaining two available colours will the system converge to a proper colouring, otherwise the non-permanent vertices will cause one or more of the permanent vertices to be unsatisfied and so to exit the permanent state at the next reset time.

With the foregoing in mind, let $\mathcal{S} := \{S_\tau : \tau = 1, 2, \ldots\}$ denote the set of the reset times when the FCFL algorithm allows vertices to exit the permanent state, and $\bar{\mathcal{S}} := \mathbb{N} \setminus \mathcal{S}$ denote all other times. What we have is that at times $t \in \bar{\mathcal{S}}$ the drift $\mathbb{E}[Z_{t+1} | Z_t] - Z_t$ must be non-positive (vertices can enter the permanent state but cannot leave it) but at times $t \in \mathcal{S}$ the drift may be positive *i.e.* $\mathbb{E}[Z_{t+1} | Z_t]$ may increase. The situation is illustrated schematically in Figure 2. To show convergence of $\mathbb{E}[Z_t]$ what we would like to show is that $\mathbb{E}[Z_t]$ decreases monotonically at the sequence of reset times $t \in \mathcal{S}$, as indicated by the dots on the dashed line in Figure 2. But of course we want more than to just show convergence, we want to show *fast* convergence and this requires tight control of the upper bound indicated by the dashed line so that it decreases sufficiently quickly.

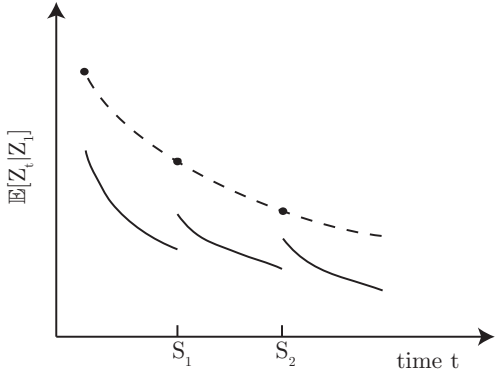

Fig. 2: Illustrating evolution of $\mathbb{E}[Z_t|Z_1]$ vs time $t$. At times $t \in \mathcal{S}$ the drift may be positive and the value may increase compared to time $t-1 \in \bar{\mathcal{S}}$. Nevertheless, at the subsequence of times $t \in \mathcal{S}$ the net effect is for $\mathbb{E}[Z_t|Z_1]$ to decrease as indicated by the dots on the dashed line.

*E. Main Result – Fast colouring with $\Delta + 1$ colours*

We present the proof in the next section, but using the approach outlined above we can show that if at least $\Delta + 1$ colours are available (where $\Delta$ is the maximum degree of the graph), FCFL is provably fast. That is it converges to a proper colouring in $\mathcal{O}(N \log N)$ time with high probability for any graph, and in $\mathcal{O}(\log N)$ time for graphs of small maximum degree, *i.e.* graphs where $\Delta = \mathcal{O}(1)$. Moreover, this is achieved while keeping the interval $S_{\tau+1} - S_\tau$ between reset times small ($\Delta + 1$), allowing the algorithm to respond quickly to topology changes.

**Theorem 2** (Fast Convergence). *Consider a CP on a graph $G = \{\mathcal{N}, \mathcal{E}\}$ with maximum degree $\Delta$ and suppose that we have $D > \Delta + 1$ available colours. Let $N := |\mathcal{N}| \geq 2$ and $\mathcal{Z}_t$ be the set of vertices in the non-permanent state at time $t \in \{0, 1, 2, \dots\}$, with $|\mathcal{Z}_t| = Z_t \in \{0, 1, 2, \dots, N\}$. Let $\tau^*$ index the first time $S_{\tau^*}$ at which a proper colouring is found. For FCFL with $S_{\tau+1} - S_\tau \geq \Delta + 1$ and $b = 1$ we have*

$$\mathbb{P}\big(\tau^* \geq B(N, \Delta, \epsilon)\big) \leq \epsilon,$$

*where*

$$B(N, \Delta, \epsilon) := \frac{\log N + \log(\epsilon^{-1}) + K}{(\Delta + 1) \log \frac{\Delta+1}{\Delta} + \frac{K}{\Delta+1}}$$

*and $K = \log \frac{1}{1+2\log 2}$.*

Observe that Theorem 2 states the convergence rate in terms of the reset times $S_\tau$. When we have $S_{\tau+1} - S_\tau \leq M$ then we can immediately express the convergence rate in terms of time slots.

**Corollary 2.** Let $R \in \{1, 2, \cdots\}$ be the earliest time such that graph $G$ is properly coloured. When $\Delta + 1 \leq S_{\tau+1} - S_\tau \leq M$, $\tau = 1, 2, \cdots$ and $S_1 \leq M$, $b = 1$ then for FCFL

$$\mathbb{P}\big(R \geq M \log N + o(1)\big) \leq \epsilon, \quad \text{as } N \to \infty.$$

*Proof:* Since $S_{\tau+1} - S_\tau \leq M$ for all $t = 1, 2, \cdots$ and $S_1 \leq M$, we can bound the convergence rate in terms of time slots, namely $MB(N, \Delta, \epsilon)$. Taking the limit of $MB(N, \Delta, \epsilon)$ as $N, \Delta \to \infty$ ends the proof. ∎

**Corollary 3** (Small degree). When $S_\tau = \tau(\Delta + 1)$, $\tau = 1, 2, \cdots$ (periodic reset times), $b = 1$ and $\Delta = \mathcal{O}(1)$, then for FCFL the convergence time $R = \mathcal{O}(\log N)$ as $N \to \infty$.

*Proof:* The thesis follows as for Corollary 2, by setting $M = \Delta + 1$ and assuming $\Delta \leq \beta$. ∎

**Corollary 4** (Almost complete graphs). When $S_\tau = \tau(\Delta+1)$, $\tau = 1, 2, \cdots$, $b = 1$ and $\Delta = \Theta(N)$, then for FCFL the convergence time $R = \mathcal{O}(N \log N)$ as $N \to \infty$.

*Proof:* The thesis follows as for Corollary 2, by setting $M = \Delta + 1$ and assuming $\lim_{N, \Delta \to \infty} \frac{\Delta}{N} = \beta > 0$,. ∎

A convergence time of $\mathcal{O}(N \log N)$ with $\Delta + 1$ colours is surprisingly close to that of the state-of-the-art given the constraints imposed by the decentralised nature of the FCFL algorithm. Faster convergence has recently been demonstrated ($\mathcal{O}(\Delta) + \frac{1}{2} \log^* N$ [4]), but only for algorithms which (i) start with a much larger number of colours and then proceed to prune these until only $\Delta + 1$ are used, (ii) require knowledge of the graph topology and (iii) make extensive use of message passing. Recall that FCFL imposes none of these requirements, and is therefore much better suited to networking applications. In Szegedy and Vishwanathan [21] it is argued that no locally-iterative $(\Delta + 1)$-colouring algorithm is likely to terminate in less than $\Omega(\Delta \log \Delta)$ rounds, implying that the FCFL algorithm may in fact be order optimal amongst locally-iterative algorithms in the case of complete graphs (when $\Delta = N$).

We note that Theorem 2 requires parameters $S_{\tau+1} - S_\tau \geq \Delta + 1$ and $b = 1$ in the FCFL algorithm. We discuss the analytic difficulty which is the source of this requirement in more detail in the next section, but this is circumvented by the introduction of the permanent state and reset times in FCFL. As already noted, the permanent state introduces memory into the FCFL algorithm which bears qualitative similarities to that introduced in CFL by selecting parameter $b$ to have a small value (typically 0.1 in [8]). This is of interest in its own right, quite apart from the resulting provably fast convergence, as it significantly broadens the class of known decentralised algorithms for graph colouring.

Further, this new approach is particularly amenable to highly efficient implementation, making it suited to use of resource constrained hardware such as RFID tags and sensors. For example, selecting a constant reset time $S_\tau = \tau(\Delta + 1)$ and uniform selection ($b = 1$) yields the following simplified instance of FCFL:

**Algorithm 2** Simplified Fast Communication-Free Learning
1: Initialise counter $t = 0$
2: Select a colour uniformly at random
3: **repeat**
4:    **if** $t = 0$ **then**
5:       $t = \Delta + 1$, $m = 0$   ▷ Reset, exit permanent state
6:    **end if**
7:    **if** $m = 0$ **then**
8:       **if** Satisfied **then**
9:          $m = 1$   ▷ Enter permanent state
10:         Leave colour unchanged.
11:      **else**
12:         Select a colour uniformly at random
13:      **end if**
14:    **else**
15:       Leave colour unchanged.
16:    **end if**
17:    $t = t - 1$
18: **until** Forever

Observe that this simplified FCFL algorithm involves no floating point arithmetic, no multiplications or divisions and only needs the availability of a uniform random number generator (which can be efficiently implemented in pseudo-random form).

## IV. ANALYSING CONVERGENCE RATE

To proceed, for each vertex $i \in \mathcal{N}$ and each time $t = 1, 2, \ldots$ define the random variable,

$$X_i(t) = \begin{cases} 1, & \text{if vertex } i \text{ is permanent at time } t \\ 0, & \text{otherwise.} \end{cases}$$

Letting $\mathcal{Z}_t = \{i : X_i(t) = 0\}$ denote the set of non-permanent vertices at time $t$ then $Z_t = |\mathcal{Z}_t|$, the cardinality of $\mathcal{Z}_t$. Now,

$$\mathbb{E}[Z_{t+1}|Z_t = Z] = \sum_{\substack{\mathcal{Z}:|\mathcal{Z}|=Z,\\ \mathcal{Z}\subset\mathcal{N}}} (\Phi_t(\mathcal{Z}) + \Psi_t(\mathcal{Z}))\, \mathbb{P}(\mathcal{Z}|Z_t = Z),$$

where

$$\Phi_t(\mathcal{Z}) := \mathbb{E}\left[\sum_{i\in\mathcal{Z}}(1 - X_i(t+1))\,\Big|\,\mathcal{Z}_t = \mathcal{Z}\right]$$

$$\Psi_t(\mathcal{Z}) := \mathbb{E}\left[\sum_{i\in\mathcal{N}\setminus\mathcal{Z}_t}(1 - X_i(t+1))\,\Big|\,\mathcal{Z}_t = \mathcal{Z}\right].$$

Suppose, for now, that we have bounds $\Phi_t(\mathcal{Z}) \leq \phi_t(Z)$, $\Psi_t(\mathcal{Z}) \leq \psi_t(Z)$ where $Z = |\mathcal{Z}|$. That is, $\Phi_t(\mathcal{Z})$ and $\Psi_t(\mathcal{Z})$ can be upper bounded by functions which depend only on the cardinality of set $\mathcal{Z}$. Then it follows that,

$$\mathbb{E}[Z_{t+1}|Z_t = Z] \leq \phi_t(Z_t) + \psi_t(Z_t).$$

Recall $\mathcal{S} := \{S_\tau : \tau = 1, 2, \ldots\}$ is the set of the reset times when the FCFL algorithm allows vertices to exit the permanent state, and $\bar{\mathcal{S}} := \mathbb{N} \setminus \mathcal{S}$. For slots $t \in \bar{\mathcal{S}}$ vertices cannot exit the permanent state and so $\psi_t(Z) = 0$. Hence,

$$\mathbb{E}[Z_{t+1}|Z_t] \leq \begin{cases} \phi_t(Z_t) & t \in \bar{\mathcal{S}} \\ \phi_t(Z_t) + \psi_t(Z_t) & t \in \mathcal{S}. \end{cases}$$

In order to streamline the discussion, and because they are satisfied by the FCFL algorithm, we make the following assumptions: (i) $\phi_t(\cdot)$ is linear and time-invariant *i.e.* $\phi_t(Z) = aZ$ with $a \geq 0$ and (ii) $\psi_t(\cdot)$ is concave. For slots $t \in \bar{\mathcal{S}}$ we then have,

$$\mathbb{E}[Z_{t+1}] = \mathbb{E}\big[\mathbb{E}[Z_{t+1}|Z_t]\big] \leq \mathbb{E}[\phi_t(Z_t)] \stackrel{(a)}{=} \phi_t(\mathbb{E}[Z_t])$$

where (a) follows since $\phi_t(\cdot)$ is linear. Hence,

$$\mathbb{E}[Z_2] \leq \phi_1(\mathbb{E}[Z_1])$$
$$\mathbb{E}[Z_3] \leq \phi_2(\mathbb{E}[Z_2]) \leq \phi_2(\phi_1(\mathbb{E}[Z_1]))$$

and

$$\mathbb{E}[Z_{S_1}] \leq \phi^{(S_1 - 1)}(\mathbb{E}[Z_1]), \tag{4}$$

where $\phi^{(t)}(Z) := \phi_t \circ \cdots \circ \phi_1(Z)$ and $\circ$ denotes function composition *i.e.* $\phi_{t+1} \circ \phi_t(Z) = \phi_{t+1}(\phi_t(Z))$. This simplifies to $\phi^{(t)}(Z) \leq \alpha^t Z$ under the assumption that $\phi(\cdot)$ is linear. Now,

$$\mathbb{E}[Z_{S_1+1}] \leq \mathbb{E}[\phi(Z_{S_1})] + \mathbb{E}[\psi_{S_1}(Z_{S_1})]$$
$$\stackrel{(b)}{\leq} \phi(\mathbb{E}[Z_{S_1}]) + \psi_{S_1}(\mathbb{E}[Z_{S_1}])$$
$$\leq \alpha^{S_1} \mathbb{E}[Z_1] + \psi_{S_1}(\alpha^{S_1 - 1} \mathbb{E}[Z_1]),$$

where (b) follows since $\psi(\cdot)$ is concave. Hence, once again using the assumption that $\phi(\cdot)$ is linear,

$$\mathbb{E}[Z_{S_2+1}] \leq \alpha^{S_2} \mathbb{E}[Z_1] + \psi_{S_2}(\alpha^{S_2 - S_1 - 1} \psi_{S_1}(\alpha^{S_1 - 1} \mathbb{E}[Z_1])), \tag{5}$$

and we can repeat to obtain $\mathbb{E}[Z_{S_\tau + 1}]$ for $\tau = 1, 2, \ldots$. However, it can be seen that the term involving $\psi_t(\cdot)$ in the expression for $\mathbb{E}[Z_{S_\tau + 1}]$ quickly becomes complex and messy. It is precisely this term which captures the positive drift at times $t \in \mathcal{S}$ and which, as noted previously, makes the analysis tricky. The key insight underlying our analysis here is that in the case of the FCFL algorithm this term can be successfully controlled via a upper bound which is tractable yet tight enough to allow fast convergence to be established.

### A. Bounding $\Phi_t(\mathcal{Z})$ for FCFL

Recall that $\Phi_t(\mathcal{Z})$ is the expected number of vertices that remain in the non-permanent state at time $t$ conditioned on the set of non-permanent vertices being $\mathcal{Z}$. The probability that a vertex leaves the non-permanent state is characterised by the following lemma:

**Lemma 2** (Entering Permanent State)**.** We have that $\mathbb{P}(X_i(t+1) = 1 | X_i(t) = 0)$ is the probability that a vertex $i$ which is in the non-permanent state at time $t$ enters the permanent state at time $t + 1$. When the number of colours $D \geq \Delta + 1$ then,

$$\mathbb{P}(X_i(t+1) = 1 | X_i(t) = 0) \geq \alpha := \frac{b}{\Delta + 1}.$$

*Proof:* A non-permanent vertex has at least $D - \Delta$ available colours, and its choice is uniform, so it has a probability at least equal to $b\frac{(D-\Delta)}{D}$ to choose a colour not used by any neighbour. Now $\frac{\Delta}{D} \leq \frac{\Delta}{\Delta+1}$, because $D \geq \Delta + 1$; so we have $b\frac{(D-\Delta)}{D} = b(1 - \frac{\Delta}{D}) \geq b(1 - \frac{\Delta}{\Delta+1}) = \frac{b}{\Delta+1}$. ∎

Lemma 2 is intuitive. By design, in FCFL the probability that a non-permanent vertex selects a colour is at least $b/D$ *i.e.* every colour has a uniformly lower bounded chance of being selected. When the number of colours $D \geq \Delta + 1$ then there is always at least one choice of colour different from that of every neighbour (since there can be at most $\Delta$ neighbours). Selecting this colour will cause the vertex to become satisfied and enter the permanent state. Note that when $b = 1$ this bound is tight, *i.e.* there exists a degree $\Delta + 1$ graph and a configuration of $D = \Delta + 1$ vertex colours for which it is satisfied with equality.

It follows from Lemma 2 that,

$$\Phi_t(\mathcal{Z}) = \mathbb{E}\left[\sum_{i \in \mathcal{Z}}(1 - X_i(t+1))|Z_t = \mathcal{Z}\right]$$
$$= Z - \mathbb{E}\left[\sum_{i \in \mathcal{Z}} X_i(t+1)|Z_t = \mathcal{Z}\right]$$
$$\leq Z - \alpha Z = \frac{\Delta + 1 - b}{\Delta + 1} Z =: \phi(Z). \quad (6)$$

Observe that $\phi(Z)$ in (6) is linear in $Z$ and time-invariant, as required.

### B. Bounding $\Psi_t(\mathcal{Z})$ for FCFL

Now we turn to $\Psi(Z)$ for the FCFL algorithm. Recall that,

$$\Psi_t(\mathcal{Z}) = \begin{cases} 0 & t \in \bar{\mathcal{S}} \\ \mathbb{E}\left[\sum_{i \in \mathcal{N} \setminus \mathcal{Z}}(1 - X_i(t+1))|\mathcal{Z}_t = \mathcal{Z}\right] & t \in \mathcal{S}. \end{cases}$$

Now $\mathcal{N} \setminus \mathcal{Z}_t$ is the set of vertices which are in a permanent state at time $t$ *i.e.* for which $X_i(t) = 1$. The following lemma bounds $\mathbb{P}(X_i(t+1) = 1|X_i(t) = 1)$ *i.e.* the probability that $X_i(t+1)$ remains equal to 1 (and so vertex $i$ stays in the permanent state) for vertices $i \in \mathcal{N} \setminus \mathcal{Z}_t$.

**Lemma 3** (Remaining Permanent). When the number of colours $D \geq \Delta + 1$ then at times $t \in \mathcal{S}$,

$$\mathbb{P}(X_i(t+1) = 1|X_i(t) = 1) \geq \left(\frac{\Delta + 1 - b}{\Delta + 1}\right)^{n(i,t)}, \quad (7)$$

where $n(i,t)$ is the number of neighbours of vertex $i$ that are in the non-permanent state at time $t$.

*Proof:* Let $x_i$ be the colour of vertex $i$. When $t \in \mathcal{S}$, permanent vertex $i$ will still keep the same colour $x_i$. By Corollary 1, other permanent vertices cannot affect the satisfaction of vertex $i$, but $i$ could lose its (permanent) state if at least one of its non-permanent neighbours chooses $x_i$. The probability that a non-permanent neighbour chooses a different colour from $x_i$ is $1 - \frac{b}{D}$ (line 14 of Algorithm 1), and since the choice of each vertex is independent, the probability all non-permanent vertices choose a different colour from $x_i$ is

$(1 - b/D)^{n(i,t)}$. Now, since $D \geq \Delta + 1$, we have $\frac{b}{D} \leq \frac{b}{\Delta+1}$ and so $1 - \frac{b}{D} \geq 1 - \frac{b}{\Delta+1} = \frac{\Delta+1-b}{\Delta+1}$. ∎

Observe that, once again, when $b = 1$ the bound in Lemma 3 is tight. It follows from Lemma 3 that

$$\Psi_t(\mathcal{Z}) \leq \sum_{i \in \mathcal{N} \setminus \mathcal{Z}_t} \left(1 - \left(\frac{\Delta + 1 - b}{\Delta + 1}\right)^{n(i,t)}\right), \; t \in \mathcal{S}. \quad (8)$$

While this provides an upper bound on $\Psi_t(\mathcal{Z})$, this bound is difficult to evaluate since it depends on $n(i,t)$, the number of neighbours of vertex $i$ that are in a non-permanent state at time $t$. We could try to use fact that $n(i,t) \leq \Delta$ to simplify this bound to,

$$\sum_{i \in \mathcal{N} \setminus \mathcal{Z}_t} \left(1 - \left(\frac{\Delta + 1 - b}{\Delta + 1}\right)^{n(i,t)}\right)$$
$$\leq \left(1 - \left(\frac{\Delta + 1 - b}{\Delta + 1}\right)^{\Delta}\right)(N - Z). \quad (9)$$

Unfortunately, however, it turns out that this upper bound is too loose to allow us to establish that $\mathbb{E}[Z_{S+1}|Z_1]$ decreases at the sequence of times $t \in \mathcal{S}$ (Figure 2): as $Z$ becomes small, $N - Z$ increases and we overestimate the number of edges affecting a vertex in the permanent state. A more sophisticated approach is needed.

We proceed as follows. Each non-permanent vertex can affect at most $\Delta$ permanent vertices (since the degree $\Delta$ is the maximum number of neighbours that a vertex can have), and by Corollary 1 a permanent vertex cannot affect any other permanent vertex. Hence, the set $\mathcal{N} \setminus \mathcal{Z}_t$ can be affected by at most a number of edges equal to

$$\sum_{i \in \mathcal{N} \setminus \mathcal{Z}_t} n(i,t) \leq \Delta Z_t, \quad (10)$$

when $\mathcal{Z}_t \neq \mathcal{N}$, and of course we must have $n(i,t) = 0$ for all $i \in \mathcal{N}$ when $\mathcal{Z}_t = \mathcal{N}$. We also have the constraints that $0 \leq n(i,t) \leq \Delta$, but it turns out that constraint (10) is sufficient for our purposes. To obtain a tighter bound on $\Psi_t(\mathcal{Z})$ we find the $n(i,t)$, $i \in \mathcal{N} \setminus \mathcal{Z}_t$ that maximise the RHS of (8) subject to the constraint (10). To do this we exploit that fact that $1 - \left(\frac{\Delta+1-b}{\Delta+1}\right)^n$ is concave in $n$ (as can be verified by inspection of the second derivative) and so the RHS of (8) is jointly convex in the $n(i,t)$, $i \in \mathcal{N} \setminus \mathcal{Z}_t$. Using this we obtain the following:

**Lemma 4** (Tighter Bound). Let $\mathcal{N} = \{1, \cdots, N\}$ and $\mathcal{Z} \subset \mathcal{N}$ be two integer sets of cardinality $N$ and $Z$ respectively, with $N > 1$ and $Z \leq N$. Let $\Delta > 1$ be an integer and $\mathbf{n} \in \mathbb{Z}^N$ a integer vector of length $N$. Suppose $\mathbf{n} = \mathbf{0}$ when $\mathcal{Z} = \mathcal{N}$ and otherwise

$$\sum_{i \in \mathcal{N} \setminus \mathcal{Z}} n_i \leq \Delta Z, \quad (11)$$

then we have that

$$\sum_{i \in \mathcal{N} \setminus \mathcal{Z}} \left(1 - \left(\frac{\Delta + 1 - b}{\Delta + 1}\right)^{n_i}\right) \leq \left(1 - \left(\frac{\Delta+1-b}{\Delta+1}\right)^{\frac{\Delta Z}{N-Z}}\right)(N - Z). \quad (12)$$

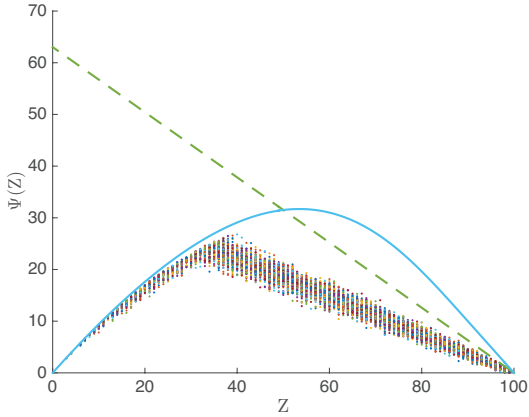

Fig. 3: Illustrating bounds on $\Psi_t(\mathcal{Z})$ vs $Z$ for $N = 100$ vertices and $b = 1$. The dashed line indicates bound (9), the solid line bound (14) and the dots indicate values of (8) for $n(i,t)$ drawn uniformly at random from the feasible set.

*Proof:* When $\mathcal{Z} = \mathcal{N}$ the LHS of (12) is formally zero and the inequality holds trivially. Otherwise, we maximise over $\mathbf{n}$ the concave function $\sum_{i \in \mathcal{N} \setminus \mathcal{Z}} \left(1 - \left(\frac{\Delta+1-b}{\Delta+1}\right)^{n_i}\right)$ subject to linear constraint (11). Since we want an upper bound, we can work on the relaxed problem in which we allow $\mathbf{n} \in \mathbb{R}^N$, because the maximum over this wider set will be greater than or equal to the maximum over $\mathbb{Z}^N$. This relaxed optimisation is convex. The Slater condition is satisfied, because $\Delta > 1$ and $Z \geq 1$ and so the point $n_i = 0 \; \forall i$ is in the interior of the constraint set. Hence, strong duality holds. The Lagrangian is $L = -\sum_{i \in \mathcal{N} \setminus \mathcal{Z}} \left(1 - \left(\frac{\Delta+1-b}{\Delta+1}\right)^{n_i}\right) + \mu(\sum_{i \in \mathcal{N} \setminus \mathcal{Z}} n_i - \Delta Z)$ and the main KKT conditions are $-\left(\frac{\Delta+1-b}{\Delta+1}\right)^{n_i} \log \frac{\Delta+1-b}{\Delta+1} = \mu$, $i \in \mathcal{N} \setminus \mathcal{Z}$. It follows that $n_i = n_j$ for all $i,j \in \mathcal{N} \setminus \mathcal{Z}$. Further, since $0 < \left(\frac{\Delta+1-b}{\Delta+1}\right) < 1$ and the $n_i$ are finite it follows that $\mu > 0$ and so by complementary slackness constraint (11) is tight. Hence, the $n_i$ maximising $f$ are

$$n_i = \frac{\Delta Z}{N - Z}, \quad i \in \mathcal{N} \setminus \mathcal{Z}, \quad Z < N, \tag{13}$$

Substituting into $\sum_{i \in \mathcal{N} \setminus \mathcal{Z}} \left(1 - \left(\frac{\Delta+1-b}{\Delta+1}\right)^{n_i}\right)$ now yields the stated result. ∎

Using Lemma 4 it follows from (8) that,

$$\Psi_t(\mathcal{Z}) \leq \left(1 - \left(\frac{\Delta+1-b}{\Delta+1}\right)^{\frac{\Delta Z}{N-Z}}\right)(N-Z) =: \psi(Z). \tag{14}$$

The bound $\psi(Z)$ depends only on the cardinality $Z$ of set $\mathcal{Z}$ and it can be verified (by inspection of the second derivative) that $\psi(Z)$ in (14) is concave in $Z$ as required. Observe also that $\psi(Z)$ is time-invariant.

The upper bound (14) is considerably tighter than (9) as $Z$ approaches 0. This can be seen, for example, in Figure 3.

### C. Some Useful Identities

The following lemma summarises identities that will prove useful in the next section.

**Lemma 5.** Let $\tilde{\Delta} := \frac{\Delta+1-b}{\Delta+1}$. For $\Delta \geq 1$ we have that:
(i) The range of $\tilde{\Delta}^{\frac{\alpha \Delta}{1-\alpha}}$ is $[e^{-\frac{\alpha}{1-\alpha}}, 1]$ for any $0 < \alpha < 1$;
(ii) $\tilde{\Delta}^{(\Delta+1)}$ is increasing in $\Delta$;
(iii) $\lim_{\Delta \to \infty} \tilde{\Delta}^{(\Delta+1)} = 1/e^b$.
(iv) $\frac{k}{e^b} < 1$ for $0 < b \leq 1$ where $k = 1 + 2\log\frac{2}{2-b}$.

*Proof:* (i) The derivative of $\tilde{\Delta}^{\frac{\alpha\Delta}{1-\alpha}}$ with respect to $\Delta$ is $\tilde{\Delta}^{\frac{\alpha\Delta}{1-\alpha}} \frac{\alpha}{1-\alpha}\left(\frac{b}{(\Delta+1)}\frac{\Delta}{\Delta+1-b} + \log \tilde{\Delta}\right) < 0$. When $\Delta \geq 1$ and $\frac{\alpha}{1-\alpha} > 0$, $0 < b \leq 1$ it follows that $\tilde{\Delta}^{\frac{\alpha\Delta}{1-\alpha}}$ is decreasing. Hence, it takes its maximum value of $\left(\frac{2-b}{2}\right)^{\frac{\alpha}{1-\alpha}} < 1$ when $\Delta = 1$ and its minimum as $\Delta \to \infty$ is $e^{-b\frac{\alpha}{1-\alpha}} \geq e^{-\frac{\alpha}{1-\alpha}}$. (ii) The derivative of $\tilde{\Delta}^{(\Delta+1)}$ with respect to $\Delta$ is $\tilde{\Delta}^{\Delta+1}\left(\frac{b}{\Delta+1-b} + \log \tilde{\Delta}\right) > 0$ when $\Delta \geq 1$. It follows that $\tilde{\Delta}^{(\Delta+1)}$ is increasing as claimed. (iii) Follows from the limit form of $e$. (iv) It can be verified by inspection of the derivative that $\frac{k}{e^b}$ is strictly decreasing on interval $0 < b \leq 1$. Hence, $\frac{k}{e^b} < \frac{1}{e^0} = 1$. ∎

### D. Bounding $\mathbb{E}[Z_t]$ As Time Elapses

As already noted, the main challenge in the analysis is controlling the $\psi_t(\cdot)$ term in (5) as time elapses. Combining (6) and (14) we have that

$$\mathbb{E}[Z_{S_\tau+1}] \leq \tilde{\Delta}\,\mathbb{E}[Z_{S_\tau}] + \left(1 - \tilde{\Delta}^{\frac{\Delta \mathbb{E}[Z_{S_\tau}]}{N-\mathbb{E}[Z_{S_\tau}]}}\right)(N - \mathbb{E}[Z_{S_\tau}]), \tag{15}$$

where $\tilde{\Delta} := \frac{\Delta+1-b}{\Delta+1}$.

To proceed we would like to substitute in (15) an upper bound on $\mathbb{E}[Z_{S_\tau}]$ rather than using the exact value. However, for the inequality in (15) to continue to hold after this substitution requires that the RHS of (15) is monotonically increasing in $\mathbb{E}[Z_{S_\tau}]$. The first term on the RHS of (15) is linear and increasing since $\Delta > 0$ and $1 - b \geq 0$. The following lemma establishes that the second term on the RHS of (15) is also increasing.

**Lemma 6** (Increasing)**.** Let $f(Z) := \left(1 - \tilde{\Delta}^{\frac{\Delta Z}{N-Z}}\right)(N-Z)$ with $\tilde{\Delta} := \frac{\Delta+1-b}{\Delta+1}$. Then $f(Z) \leq f(Y)$ whenever $0 \leq Z \leq Y \leq \alpha N$ for any $0 \leq \alpha \leq \alpha^*$ where $\alpha^*$ satisfies $\alpha^* = h(e^{-\frac{\alpha^*}{1-\alpha^*}})$ with $h(x) = \frac{x\log x}{x-1}$. Note that $\alpha = \frac{1}{2}$ is one admissible choice.

*Proof:* It can be verified by inspection of the second derivative that $f(Z)$ is concave for $Z \in [0, N]$. Hence, the supporting hyperplane property holds *i.e.* $f(Z) \leq f(Y) - (Y-Z)f'(Y)$. For $Z \leq Y$ then when $f'(Y) \geq 0$ it follows that $f(Z) \leq f(Y)$ as required. By the monotonicity of the sub gradients of concave functions $(f'(Y) - f'(Z))(Y-Z) \leq 0$. Hence, for $Y \leq \alpha N$ then $(f'(\alpha N) - f'(Y))(\alpha N - Y) \leq 0$ *i.e.* $f'(Y) \geq f'(\alpha N)$ and for $f'(Y) \geq 0$ it is sufficient to show that $f'(\alpha N) \geq 0$. Now,

$$f'(\alpha N) = -1 + \tilde{\Delta}^{\frac{\alpha\Delta}{1-\alpha}} - \frac{1}{\alpha}\tilde{\Delta}^{\frac{\alpha\Delta}{(1-\alpha)}} \log \tilde{\Delta}^{\frac{\alpha\Delta}{(1-\alpha)}}$$

The function $\tilde{f}(x) = -1 + x - \frac{1}{\alpha}x \log x$ is concave on $[0, \infty)$ (the second derivative is negative for all $x \in [0, \infty)$), and has its global maximum (equal to $-1 + \frac{1}{\alpha}e^{\alpha-1} > 0$) at $x^* = e^{\alpha-1} < 1$. It is strictly increasing on the left side and strictly decreasing on the right side of this maximum. $\tilde{f}(x)$ has one root to the right of this maximum at $x_+ = 1$ and a second root $x_-$ to the left at the point satisfying $\alpha = \frac{x_- \log x_-}{x_- - 1}$. By Lemma 5, $\tilde{\Delta}^{\frac{\alpha \Delta}{1-\alpha}}$ takes values in $[e^{-\frac{\alpha}{1-\alpha}}, 1]$. Recall that $x_+ = 1$. Observe that since $h(x) = \frac{x \log x}{x-1}$ is strictly increasing, so is its inverse $h^{-1}(\cdot)$. Hence, for $\alpha \leq h(e^{-\frac{\alpha}{1-\alpha}})$ then $x_- = h^{-1}(\alpha) \leq e^{-\frac{\alpha}{1-\alpha}}$. It follows that $\tilde{f}(x)$ is non-negative in $[e^{-\frac{\alpha}{1-\alpha}}, 1]$. It can be verified that $h(e^{-1}) > \frac{1}{2}$ and so $\alpha = \frac{1}{2}$ is an admissible choice. Observing that $f'(\alpha N) = \tilde{f}(\tilde{\Delta}^{\frac{\alpha \Delta}{\alpha-1}})$ it follows that $f'(\alpha N) \geq 0$ and we are done. ∎

Note that the condition in Lemma 6 is tight in the sense that for a graph with sufficiently large degree $\Delta$ the function $f(\cdot)$ is not increasing when $\alpha > \alpha^*$. Lemma 6 allows us to substitute an upper bound on $\mathbb{E}[Z_{S_\tau}]$ into (15). However, as we will shortly see, the resulting expression is still too complex to be manageable. To obtain a tractable expression we need to use the following Lemma.

**Lemma 7** (Clean Upper Bound). *For any $\Delta \geq 1$, $\tau \geq 1$ and $0 < b \leq 1$ we have*

$$\tilde{\Delta}^{\tau(\Delta+1)}k^{\tau-1} + \left(1 - \tilde{\Delta}^{\frac{\Delta \tilde{\Delta}^{\tau(\Delta+1)}k^{\tau-1}}{1-\tilde{\Delta}^{\tau(\Delta+1)}k^{\tau-1}}}\right)(1 - \tilde{\Delta}^{\tau(\Delta+1)}k^{\tau-1})$$
$$\leq k^\tau \tilde{\Delta}^{\tau(\Delta+1)+1}, \quad (16)$$

*where $\tilde{\Delta} := \frac{\Delta+1-b}{\Delta+1}$ and $k = 1 + 2\log\frac{2}{2-b}$.*

*Proof:* The required bound (16) can be rewritten equivalently as
$$\frac{1}{k}F(\Delta, Y) \leq 1,$$
with
$$F(\Delta, Y) = 1 + Y\frac{\Delta+1}{\Delta+1-b}\left(1 - \left(\frac{\Delta+1-b}{\Delta+1}\right)^{\Delta/Y}\right).$$
and $Y(X) := \frac{1-X}{X}$, $X(\Delta, \tau) := \left(\frac{\Delta+1-b}{\Delta+1}\right)^{\tau(\Delta+1)} k^{\tau-1}$. By Lemma 5 $\tilde{\Delta}^{\tau(\Delta+1)}$ is increasing in $\Delta$. It follows that $X(\Delta, \tau)$ is increasing in $\Delta$ since $\tau \geq 1$. Further, $X(\Delta, \tau)$ is decreasing in $\tau$ since its derivative with respect to $\tau$ is $\tilde{\Delta}^{\tau(\Delta+1)}k^{\tau-1}\log k\tilde{\Delta}^{\Delta+1} \leq e^{-b\tau}k^{\tau-1}\log k/e^b \leq 0$ as $k < e^b$ for $0 < b \leq 1$, $\tau \geq 1$ and $\lim_{\Delta \to \infty}\tilde{\Delta}^{(\Delta+1)} = 1/e^b$ (see Lemma 5). Hence, $X(\Delta, \tau)$ is bounded from above by $1/e^b$ (since $X(\Delta, \tau)$ is increasing with $\Delta$ and decreasing with $\tau$, taking the limit as $\Delta \to \infty$ we get $1/e^b$ when $\tau = 1$) and from below by 0. It follows that $Y(X) \in [e^b - 1, \infty)$. It can be verified by inspection of the derivative that $F(\Delta, Y)$ is increasing with $Y$, so we can bound it from above with $F^*(\Delta) = \lim_{Y \to \infty} F(\Delta, Y) = 1 - \log\left(\frac{\Delta+1-b}{\Delta+1}\right)^{\Delta+1}$. That is, $\frac{1}{k}F(\Delta, Y) \leq \frac{1}{k}F^*(\Delta)$. Now $F^*(\Delta)$ is decreasing with $\Delta$ and so $\frac{1}{k}F(\Delta, Y) \leq \frac{1}{k}F^*(\Delta) \leq \frac{1}{k}F^*(1) = \frac{1+2\log 2/(2-b)}{k} = 1$ for $\Delta \in \{1, 2, \ldots\}$ as required. ∎

It can be seen that the expression on the LHS of (16) is quite complicated and the upper bound on the RHS in Lemma 7, which is essential for our analysis, is not obvious. It has been obtained by considering the limiting case when $\Delta \to \infty$ and building an ansatz that is exponential decaying with $\tau$.

### E. Proof of Theorem 2

Armed with Lemmas 6 and 7 we are now in a position to prove Theorem 2. For the first $S_1 - 1$ steps, vertices in the permanent state cannot become dissatisfied, so as shown in (4) we have $\mathbb{E}[Z_{S_1}] \leq \phi^{(S_1-1)}(\mathbb{E}[Z_1])$. Since $\phi(Z)$ is strictly increasing in $Z$ we can bound $\mathbb{E}[Z_1]$ with $N$, obtaining $\mathbb{E}[Z_{S_1}] \leq \tilde{\Delta}^{S_1-1}N \leq \tilde{\Delta}^{\underline{S}-1}N$, where $\underline{S} := \min_{\tau \in \{1,2,\cdots\}} S_{\tau+1} - S_\tau \geq 0$ is the minimum interval between the $S_\tau$'s and we have used the fact that $0 < \tilde{\Delta} < 1$. When $\underline{S} \geq \Delta + 1$ we have that $\tilde{\Delta}^{\underline{S}-1} \leq \frac{1}{2}$ provided that $b = 1$. It then follows that $\mathbb{E}[Z_{S_1}] \leq N/2$ and we can use Lemma 6 to bound (15) with

$$\mathbb{E}[Z_{S_1+1}] \leq \tilde{\Delta}^{\underline{S}+1}N + \left(1 - \tilde{\Delta}^{\frac{\Delta\tilde{\Delta}^{\underline{S}}N}{N-\tilde{\Delta}^{\underline{S}}N}}\right)(N - \tilde{\Delta}^{\underline{S}}N).$$

This expression is still too complicated to be used in (4) for the next $S_2$ slots, but when $\underline{S} \geq \Delta + 1$ then we can obtain a clean upper bound using Lemma 7 with $\tau = 1$. Namely,

$$\mathbb{E}[Z_{S_1+1}] \leq kN\tilde{\Delta}^{(\Delta+1)+1} \stackrel{(a)}{\leq} \frac{k}{e^b}N,$$

where $(a)$ follows from Lemma 5. For the subsequent slots $S_1 + 1$ through $S_2 - 1$, vertices in the permanent state cannot become dissatisfied, so $\psi_t(Z) = 0$ and

$$\mathbb{E}[Z_{S_2}] \leq \tilde{\Delta}^{S_2-S_1-1}\mathbb{E}[Z_{S_1+1}]$$
$$\leq \tilde{\Delta}^{2\underline{S}-1}\mathbb{E}[Z_{S_1+1}]$$
$$\leq kN\tilde{\Delta}^{2(\Delta+1)} \leq \frac{kN}{e^{2b}}.$$

When $b \geq \frac{2}{3}$ then we have $\frac{k}{e^{2b}} < \frac{1}{2}$ and we can again apply Lemma 6 to obtain

$$\mathbb{E}[Z_{2S+1}] \leq \tilde{\Delta}^{2(\Delta+1)+1}N + \left(1 - \tilde{\Delta}^{\frac{\Delta\tilde{\Delta}^{2(\Delta+1)}N}{N-\tilde{\Delta}^{2(\Delta+1)}N}}\right)(N - \tilde{\Delta}^{2(\Delta+1)}N).$$

Now we can apply Lemma 7 again with $\tau = 2$.

Iterating this procedure, for all $\tau = 1, 2, \cdots$ we obtain

$$\mathbb{E}[Z_{S_\tau}] \leq k^{\tau-1}N\tilde{\Delta}^{\tau(\Delta+1)}. \quad (17)$$

With bound (17) we are now almost done. By Lemma 5 we have $\tilde{\Delta}^{(\Delta+1)} \leq e^{-b}$ and so the RHS of (17) is upper bounded by $\frac{N}{k}(\frac{k}{e^b})^\tau$. Since $\frac{k}{e^b} < 1$ (by Lemma 5) then (17) establishes that $\mathbb{E}[Z_t]$ is decreasing at the sequence of times $t \in \mathcal{S}$.

Recalling that $\mathbb{P}(Z_{S_\tau} \geq 1) \leq \mathbb{E}[Z_{S_\tau}]$, to ensure $\mathbb{P}(Z_{S_\tau} \geq 1) \leq \epsilon$ it follows from (17) that it is enough to choose

$$\tau \geq B(N, \Delta, \epsilon), \quad (18)$$

where $B(N, \Delta, \epsilon) := \frac{\log N + \log(\epsilon^{-1}) + \log(k^{-1})}{(\Delta+1)\log(\frac{\Delta+1}{\Delta}) + \log(k^{-1})}$, from which Theorem 2 now follows.

Bound (18) gives the convergence rate in terms of the reset times $S_\tau$. When we have $S_{\tau+1} - S_\tau \leq M$ for all $t = 1, 2, \cdots$ and $S_1 \leq M$ then we can immediately use this to bound the convergence rate in terms of time slots, namely $MB(N, \Delta, \epsilon)$.

Taking the limit of $M\frac{\log N+\log(\epsilon^{-1})+\log(k^{-1})}{1+\log(k^{-1})}$ as $N,\Delta \to \infty$ now yields Corollary 2. Corollary 3 is obtained setting $M = \Delta+1$ and assuming $\Delta \leq \beta$. Corollary 4 is obtained by setting $M = \Delta + 1$ and assuming $\lim_{N,\Delta \to \infty} \frac{\Delta}{N} = \beta > 0$.

*F. Discussion*

In this analysis we have taken care to ensure that the bounds used are tight *i.e.* there exists a graph and an assignment of colours for which they are satisfied with equality. This suggests that the bound on convergence rate in Theorem 2 is probably almost as good as we can do without restricting attention to specific types of graph.

The requirement in Theorem 2 that $S_{\tau+1} - S_\tau \geq \Delta+1$ arises from Lemma 7, until that point there is no restriction on the choice of the reset times $S_\tau$. Extending the analysis to settings where $S_{\tau+1} - S_\tau < \Delta+1$ therefore requires extending Lemma 7. However, obtaining Lemma 7 was already a difficult step and its extension likely requires development of a new analysis approach *e.g.* a new stochastic concentration bound.

The requirement that $b = 1$ arises from application of Lemma 6, until this point there is no restriction on the value of parameter $b$. Lemma 6 is used to ensure that the RHS of (15) is increasing and so we can substitute an upper bound for $\mathbb{E}[Z_{S_t}]$ while preserving the inequality. Lemma 6 itself gives an exact condition. However, it might be possible to relax the requirement on $b$ by allowing the RHS of (15) to decrease in a controlled way and modifying the inequality in (15) after the substitution of the bound for $\mathbb{E}[Z_{S_t}]$ accordingly. However, this also has knock-on effects on the application of Lemma 7, so we leave it as future work. An alternative is to relax the requirement that $\underline{S} \geq \Delta+1$ to one that $\underline{S} \geq \Delta+n$ with $n > 1$. For example, selecting $n = 2$ ensures $\tilde{\Delta}^{\underline{S}-1} = \tilde{\Delta}^{\Delta+1} \leq 1/e^b$ which is less than $1/2$ for $b > \log(2) \approx 0.69$.

## V. NUMERICAL SIMULATIONS

*A. Convergence Rate*

Theorem 2 only provides an upper bound on the convergence rate. In Figure 4 we compare the measured convergence time with this bound for a range of graph types (bipartite, complete, 12-partite) and sizes (up to $N = 2000$ vertices), over $10\,000$ runs of the algorithm. Figure 4 plots the ratio between these, which can be seen to tend towards a constant value as $N$ increases so confirming that the $\mathcal{O}((\Delta+1)\log N)$ behaviour in Theorem 2 indeed broadly captures the actual scaling of convergence time with $N$. The ratio in Figure 4 is, however, less than one which indicates that there may be scope to further refine the prefactor in the Theorem 2 bound, at least for specific classes of graphs. We note that similar results are obtained for random graphs, although we do not include them to save space.

*B. Behaviour on Graph Change*

We analyse the case in which the algorithm has already converged to a proper coloring and then a number of vertices in the graph change change color. In Figure 5, the convergence time after a perturbation of the 2% of the vertices is shown and compared to the convergence time starting from a state when all vertices are dissatisfied. The results are for complete graphs of 60 vertices with 20% of the edges removed. This is a challenging case since perturbations to the color of a vertex propagate quickly to affect many neighbours. For comparison, the corresponding measured convergence rate of the Learning-BEB algorithm (that has exponential behaviour) has also been computed. For FCFL data is shown for 1000 runs of the algorithm, but for Learning-BEB we used fewer runs as each run was extremely slow to complete being around 6 orders of magnitude longer than with FCFL even when only 2% of the nodes are perturbed, and thus are not shown in the figure. This example highlights the importance of fast convergence since with decentralised algorithms even a small perturbation can easily disrupt a strongly connected graph.

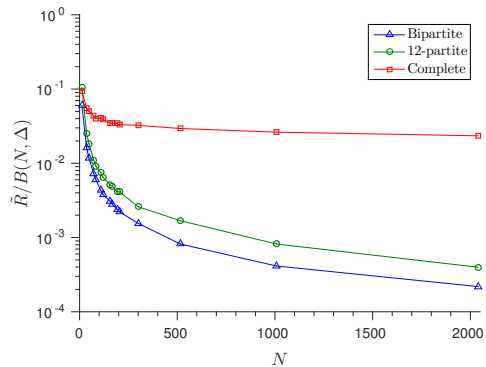

Fig. 4: Ratio between measured time for simplified FCFL (Algorithm 2) to converge and the bound $(\Delta+1)B(N,\Delta,1/2)$. Median over $10\,000$ runs of FCFL with parameters $S_\tau = \tau(\Delta+1)$, $b = 1$.

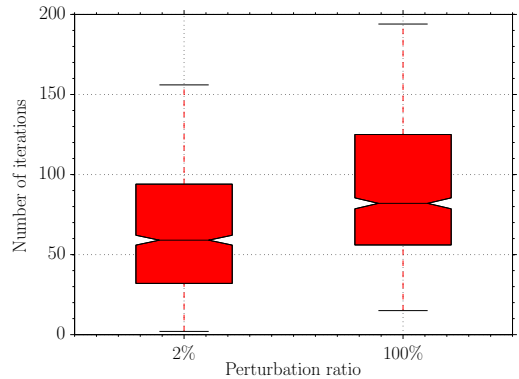

Fig. 5: Convergence time of FCFL for already correctly colored graph after a perturbation.

We note that deriving analytic convergence rate bounds for cases in which the rate of change of the graph is comparable to the convergence rate of the algorithm is a substantial undertaking in its own right and out of the scope of the present paper.

## C. Brief Application Example: Reading RFID Tags

We illustrate application of the FCFL to avoid collisions when reading RFID tags. Communication from the tags to the reader can fail when there is a collision, *i. e.* when at least two RFID tags within the coverage of the reader transmit at the same time. To mitigate this, the RFID protocol implements a basic slotted Aloha collision resolution mechanism [10, 19, 22].

When the reader needs to identify a tag, it issues a QUERY command, and each tag in the coverage area selects an integer u.a.r. in interval $[0, D-1]$, where parameter $D-1$ is set by the reader. All tags that select 0 reply immediately; tags that select another number record those numbers in a counter and don't transmit. A tag replies by sending a 16 bit random number. If the reader hears the random number, it echoes that number back as an acknowledgement, causing the tag to send its Electronic Product Code (EPC). The reader can then send commands specific to that tag, or continue to inventory other tags. In case of collision or the need for another identification, the reader can issue a QUERY REP command, causing all of the tags to decrement their counters by 1; again, any tag reaching a counter value of 0 will respond. After $M$ steps, the procedure can start again with a QUERY command. The reader can set a flag (flag B) on successfully read tags, so they will not answer anymore to subsequent queries until the tag reverts to flag A (usually after a time between 500 ms and 5 s, but no upper limit is set in the protocol).

Our aim is to implement a collision resolution mechanism that possesses the following properties: (i) allows tags to be detected quickly (reading time comparable with Aloha); (ii) allows subsequent reads per tag to be faster; (iii) allows the reader to correctly read all of the tags when their relative positions change; (iv) the new mechanism is backward compatible *i. e.* able to work with both standard RFID tags and new tags.

The task of assigning a different time slot (different counter value when the QUERY command is issued) to each RFID tag can be mapped to a CP on a graph, where the structure of the graph depends on the location of the tags. Namely, graph $G = (\mathcal{N}, \mathcal{E})$ is built such that $\mathcal{N}$ is the set of tags, and an edge $e = (i, j) \in \mathcal{E}$ iff the tags $i$ and $j$ are near enough for their transmissions to potentially collide. When all tags are within the coverage range of the reader, the problem is mapped to colouring of a complete graph. More generally (*e. g.* when the reader can cover at most $k$ tags per time), many RFID applications can be modeled as a CP on a complete $k$-partite graph $G_{s_1,\ldots,s_k}$, *i. e.* the graph composed of $k$ independent sets of (possibly different) size $s_i, i = 1, \ldots, k$, such that each set is connected with all the vertices of the other sets. This graph is $k$-colourable.

The FCFL algorithm can be implemented in an existing RFID infrastructure, ensuring backward compatibility, as follows. The idea is to modify the behaviour of the tag to allow it to enter the permanent state after a successful QUERY, and to possibly exit it every $\underline{S}$ periods by extending the meaning of the *QueryAdjust* command. The QueryAdjust command is normally used to modify the range $[0, D-1]$ in the tags, to reduce the collision probability when many tags are present, or to reduce the expected backoff when few of them are present. A way to implement the reset capability is to let the tag exit the permanent state when a QueryAdjust command is received[3]. Modified tags will thus have the following additional capabilities: (i) if the reader sets flag B the tag will enter the *permanent* state, keeping in memory the random number that allowed the communication, and stop answering to queries, (ii) if the tag receives a *QueryAdjust* broadcast command it will exit the permanent state by reverting to flag A and thus becoming ready to decrement the counter when the QUERY REP command is broadcast again. As for the QUERY command, if the stored counter is equal to 0, the tag will immediately attempt transmission. The reader is programmed to send a *QueryAdjust* command every $\underline{S}$ period, *i. e.* every $\underline{S} \cdot D$ queries, immediately after the QUERY command, so that the tags that are leaving the permanent state will still select the same random number, to potentially re-enter immediately the permanent state, if no tag is colliding with them. The reader also sets flag B on each tag that is correctly detected *in a time slot not used by previously detected tags*. In this way already identified tags will not cause collisions, and tags that are correctly identified but that would cause a collision (with a previous identified tag) will continue to change.

This implementation will still work together with non-modified tags at the expense of having some collisions, because those non-modified tags will choose a new (possibly different) time slot at every new QUERY, but each non modified tag can at most affect one modified tag, so the overall performance should still be superior to the standard slotted Aloha mechanism.

We compare the convergence time of the FCFL algorithm to each of the following algorithms from [16]:

**BFSA** Basic Framed Slotted Aloha, with standard superframe size of $D = 256$ slots.

**DFSA** Dynamic Framed Slotted Aloha, where the superframe size $D$ doubles when the number of slots with collisions is larger than 70 % of the current superframe size, and halves when the number of slots with collisions is less than 30 %.

**EDFSA** Enhanced Dynamic Framed Slotted Aloha, see [16] for more details of this enhanced version of DFSA.

For these algorithms the superframe size $D$ is the number of slots after which the reader starts a new QUERY (forcing the tags to select a new slot u.a.r.). These algorithms are all *memoryless* in the sense that over each superframe they behave statistically in the same way. In contrast, the FCFL algorithm has a transient period during which a collision-free schedule is determined and after which tags will deterministically select the same slot at every subsequent superframe in a collision-free manner.

Measurements of time taken to read all tags are given in Table I for these algorithms. It can be seen that the FCFL algorithm is comparable with classic slotted Aloha during the transient period, but once in steady state performs considerably better (yielding a 83 % reduction in read time), and also better than the state-of-the-art dynamically adjusted slotted Aloha (over which FCFL offers a $66\,\%$ reduction in read time). Using the ISO15693 high tag data rate [22], the reader needs at each slot $1\,\mathrm{ms}$ to send the QUERY (or QUERY REP) command, and the tag needs $6\,\mathrm{ms}$ to complete the identification procedure with the reader (for transmission of the random number and reception of the echo acknowledgement). This would mean that FCFL allows 1000 tags to be read in around 7 seconds compared with more than 40 seconds for classic slotted Aloha.

---

[3]The *QueryAdjust* command will still be able to set the new value of $D$ as in the original command, the only change is the behaviour of the tag.

**Algorithm 3** Simplified Fast Communication-Free Learning for RFID
___
　　Reader block
1: Broadcast QueryAdjust command with $D$　　　　▷ Reset
2: Initialise counter $t = 0$
3: **repeat**
4: 　**if** $t = 0 \mod D$ **then**
5: 　　Broadcast QUERY command
6: 　　**if** $t = 0 \mod \underline{S}D$ **then**
7: 　　　Broadcast QueryAdjust command with $D$
8: 　　**end if**
9: 　**else**
10: 　　Broadcast QUERY REP command
11: 　　**if** New tag $T$ detected in an unused slot **then**
12: 　　　Add $T$ to inventory　　▷ (and any additional operation)
13: 　　　Set Flag=B to tag $T$
14: 　　**end if**
15: 　**end if**
16: 　t=t+1
17: **until** Forever
　　Tag block
18: **if** Flag=A **then**
19: 　**if** $C = 0$ **then**
20: 　　Send EPC to reader and establish connection
21: 　**end if**
22: 　**if** Received QUERY command **then**
23: 　　Select an integer $C$ in $[0, D-1]$
24: 　**end if**
25: 　**if** Received QUERY REP **then**
26: 　　$C = C - 1$
27: 　**end if**
28: **else**　　　　　　　　　　　　　　　　　▷ Flag=B
29: 　**if** Received QueryAdjust **then**
30: 　　Set Flag=A, $C = C - 1$
31: 　**end if**
32: **end if**
___

|  | 200 tags | | 1000 tags | |
|---|---|---|---|---|
| Algorithm | First inventory | At steady state | First inventory | At steady state |
| BFSA | 1280 | 1280 | 5850 | 5850 |
| DFSA | 662 | 662 | 5425 | 5425 |
| EDFSA | 628 | 628 | 2916 | 2916 |
| **FCFL** | 816 | **200** | 5040 | **1000** |

TABLE I: Median number of time slots needed to correctly read all tags vs the algorithm used (for FCFL we used the simplified version, *i. e.* Algorithm 2). Complete graph with $N = 200$ and $N = 1000$.

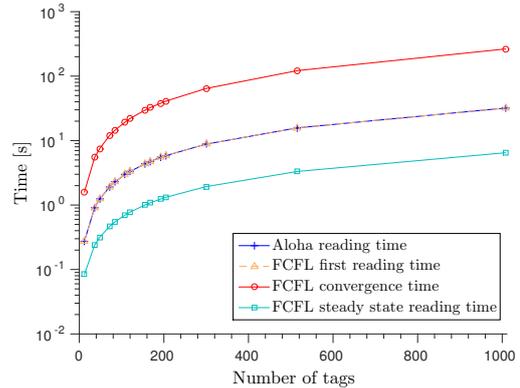

Fig. 6: Reading time of simplified FCFL (Algorithm 2) and slotted Aloha vs number of tags. 12-partite complete graph. Median over $10\,000$ runs.

Figure 6 plots the measured convergence time (median over $10\,000$ runs), the time taken to read all tags initially and the time taken in steady state. Data is shown both for FCFL and classic slotted Aloha with flagging enabled and superframe size $D$ equal to $\Delta + 1$. It can be seen that the initial read time is comparable for the both algorithms, but after convergence (less than 5 minutes for a shelf of 1000 items) the FCFL algorithm is able to check the status of all tags within 7 seconds, compared with the 32 seconds required by Aloha, a time saving of around $450\%$.

## VI. CONCLUSIONS AND FUTURE WORK

In this paper we consider algorithms for quickly solving, in a fully decentralised way (*i.e.* with no message passing), the classic problem of colouring a graph. We propose a novel algorithm that is automatically responsive to topology changes, and we prove that it converges quickly to a proper colouring in $\mathcal{O}(N \log N)$ time with high probability for generic graphs (and in $\mathcal{O}(\log N)$ time if $\Delta = \mathcal{O}(1)$) when the number of available colours is greater than $\Delta$, the maximum degree of the graph. We believe the proof techniques used in this work are of independent interest and provide new insight into the properties required to ensure fast convergence of decentralised algorithms.

We note that application of FCFL to general constraint satisfaction problems is direct, but we leave analysis of convergence rate in this more general setting to future work.

### ACKNOWLEDGMENTS

We want to thank Jaume Barcelo, Joan Meli-Segu and Marc Morenza for their thoughtful insights. Their expertise on RFID and Learning-BEB substantially improved this work.

## Appendix

Consider graph $G = (\mathcal{N}, \mathcal{E})$. Let $A$ denote the set of assignments which are absorbing for FCFL algorithm, *i.e.* the set of proper colourings. All absorbing assignments are also satisfying. When the colouring problem is feasible (the number of colours available is greater than or equal to $\chi$) then $A \neq \emptyset$ (at least one satisfying assignment exists). Let $a \in A$ be a target satisfying assignment. We will refer to the assignment at time step $t$ as $\vec{x}(t)$. Let $U_{\vec{x}(t)}$ denote the set of unsatisfied vertices and $\mathcal{D}$ the set of available colours. Define $\gamma = b/D$.

**Lemma 8.** *If a vertex is unsatisfied, when using the FCFL algorithm the probability that the vertex chooses any colour $j$ at the next step is greater than or equal to $\gamma$.*

*Proof:* This follows from step 11 of FCFL algorithm. ∎

*Proof of Theorem 1:* Consider the FCFL algorithm starting from an assignment $\vec{x}(0)$. Select an arbitrary valid solution $a \in A$. Since the CP is satisfiable, we have that $A \neq \emptyset$. We will exhibit a sequence of events that, regardless of the initial configuration, leads to a satisfying assignment with a probability for which we find a lower bound.

At the first step we consider the chain of events that changes the assignment, after $S_1$ steps, to

$$x_i(S_1+1) = \begin{cases} a_i & \text{if } i \in U_{\vec{x}(0)}, \\ x_i(0) & \text{otherwise}. \end{cases} \quad (19)$$

This is feasible since the FCFL algorithm ensures that all satisfied vertices at step 0 will remain unchanged for at least

$S_1$ steps and each unsatisfied vertex may change its colour at step 1, and keep the same colour for $S_1$ steps with probability at least $\gamma^{S_1}$ (by Lemma 8). The probability that this event happens is greater than $\gamma^{S_1|U_{\vec{x}(0)}|}$. Now, the set of unsatisfied variables could have changed. If $U_{\vec{x}(S_1+1)} = \emptyset$, we have finished, otherwise we consider again the event that changes the assignment similarly to equation (19), *i. e.* at reset time $S_\tau$ we have

$$x_i(S_\tau + 1) = \begin{cases} a_i & \text{if } i \in U_{\vec{x}(S_{\tau-1})}, \\ x_i(S_{\tau-1}) & \text{otherwise.} \end{cases}$$

The probability of this happening is greater than $\gamma^{(S_\tau - S_{\tau-1})|U_{\vec{x}(S_{\tau-1})}|}$. The lower bound on the probability of this sequence is obtained when at each reset time only one new vertex chooses the target colouring added, giving us the bound of $S_N$ steps, with probability greater than $\gamma^{M \cdot 1} \cdot \gamma^{M \cdot 2} \ldots \gamma^{M \cdot N} = \gamma^{MN(N+1)/2}$. This can be easily seen considering any other sequence where some vertices choose the target colouring at the same time: let $\mathcal{K}$ be such sequence. Clearly $\mathcal{K} \subset \{1, 2, \ldots, N\}|$ with $|\mathcal{K}| < |\{1, 2, \ldots, N\}|$ and the corresponding probability will be $\prod_{k \in \mathcal{K}} \gamma^{M \cdot k} > \gamma^{MN(N+1)/2}$.

Due to the Markovian nature of the FCFL algorithm and the independence of the probability of the above sequence on its initial conditions, if this sequence does not occur by time $S_N$, it has the same probability of occurring by time $S_{2N}$. The probability of convergence in $k \cdot MN$ steps is greater than

$$1 - \left(1 - \gamma^{\frac{MN(N+1)}{2}}\right)^k.\qquad\blacksquare$$

**Alessandro Checco** graduated from the University of Rome "Tor Vergata" in 2010 and was awarded his PhD in mathematics from the Hamilton Institute, NUI Maynooth in 2014. In 2015 he worked on recommender systems as postoctoral researcher in Trinity College, Dublin. He is currently a research associate at the Information School, University of Sheffield. His main research interests are decentralised networks, information retrieval, human computation and data privacy.

**Doug Leith** graduated from the University of Glasgow in 1986 and was awarded his PhD, also from the University of Glasgow, in 1989. In 2001, Prof. Leith moved to the National University of Ireland, Maynooth and then in Dec 2014 to Trinity College Dublin to take up the Chair of Computer Systems in the School of Computer Science and Statistics. His current research interests include wireless networks, network congestion control, distributed optimisation and data privacy.